\documentclass[preprint,amsmath,amssymb,aps,prb,superscriptaddress,10pt]{revtex4-1}

\usepackage{graphicx}
\usepackage{dcolumn}
\usepackage{bm}
\usepackage{subfigure}

\usepackage[colorlinks=true,urlcolor=blue,linkcolor=blue,citecolor=blue,bookmarks=false]{hyperref}

\begin{document}

\title{Coupling microwave photons to topological spin-textures in Cu$_2$OSeO$_3$}

\author{S. Khan}
\email{safe.khan.11@ucl.ac.uk}
\affiliation{London Centre for Nanotechnology, University College London, London, WC1H 0AH, United Kingdom}
 
\author{O. Lee}
\affiliation{London Centre for Nanotechnology, University College London, London, WC1H 0AH, United Kingdom}
 
\author{T. Dion}
\affiliation{London Centre for Nanotechnology, University College London, London, WC1H 0AH, United Kingdom}
 
\author{C. W. Zollitsch}
\affiliation{London Centre for Nanotechnology, University College London, London, WC1H 0AH, United Kingdom}
 
\author{S. Seki}
\affiliation{Department of Applied Physics, University of Tokyo, Tokyo 113-8656, Japan}

\author{Y. Tokura}
\affiliation{Department of Applied Physics, University of Tokyo, Tokyo 113-8656, Japan}
\affiliation{RIKEN Center for Emergent Matter Science (CEMS), Wako 351-0198, Japan}
\affiliation{Tokyo College, University of Tokyo, Tokyo 113-8656, Japan}

\author{J. D. Breeze}
\affiliation{Department of Materials, Imperial College London, Exhibition Road, London, SW7 2AZ, United Kingdom}
 
 \author{H. Kurebayashi}
 \email{h.kurebayashi@ucl.ac.uk}
 \affiliation{London Centre for Nanotechnology, University College London, London, WC1H 0AH, United Kingdom}

\date{\today}

\begin{abstract}
Topologically protected nanoscale spin textures, known as magnetic skyrmions, possess particle-like properties and feature emergent magnetism effects. In bulk cubic heli-magnets, distinct skyrmion resonant modes are already identified using a technique like ferromagnetic resonance in spintronics. However, direct light-matter coupling between microwave photons and skyrmion resonance modes has not been demonstrated yet. Utilising two distinct cavity systems, we realise to observe a direct interaction between the cavity resonant mode and two resonant skyrmion modes, the counter-clockwise gyration and breathing modes, in bulk Cu$_2$OSeO$_3$. For both resonant modes, we find the largest coupling strength at 57 K indicated by an enhancement of the cavity linewidth at the degeneracy point. We study the effective coupling strength as a function of temperature within the expected skyrmion phase. We attribute the maximum in effective coupling strength to the presence of a large number of skyrmions, and correspondingly to a completely stable skyrmion lattice. Our experimental findings indicate that the coupling between photons and resonant modes of magnetic skyrmions depends on the relative density of these topological particles instead of the pure spin number in the system. 

\end{abstract}

\maketitle

\section*{Introduction}
In condensed matter physics, the study of light-matter interactions holds an important stature as it provides physical insight into properties of coupled systems while facilitating the design of new-concept devices \cite{li2020hybrid, clerk2020hybrid,awschalom2021quantum}. The light-matter coupling in this context between microwave photons residing in a cavity and particle ensembles has been studied in a wide range of material systems \cite{bhoi2019photon}. These systems include two-dimensional electron gases \cite{muravev2011observation,scalari2012ultrastrong}, paramagnetic spins \cite{chiorescu2010magnetic, breeze2017room} and exchange interaction dominated ordered spin systems \cite{tabuchi2014hybridizing, zhang2014strongly, goryachev2014high,huebl2013high}. Quantum information processing as quantum memories \cite{schoelkopf2008wiring, kurizki2015quantum} and quantum transducers \cite{blum2015interfacing} are particularly interesting applications which could be made possible due to the hybrid spin-ensemble-photon systems. Formation of a quasi-particle, known as the cavity-magnon-polariton (CMP), results from the strong coherent coupling phenomenon between cavity photons and
magnons (i.e. a quanta of spin-waves) creating a hybridised state between the two \cite{harder2018cavity}. One main advantage for using ferromagnetic materials as a base for spin-photon hybrid systems is the large spin number ($N$), as this yields very sizeable collective coupling strength ($\textrm{g}_{\textrm{c}}$) between the two systems due to the scaling condition given as $\textrm{g}_{\textrm{c}} = \textrm{g}_0 \sqrt{N}$, where $\textrm{g}_0$ is the single spin - single photon coupling strength \cite{agarwal1984vacuum, imamouglu2009cavity}.

In our earlier work \cite{abdurakhimov2019magnon}, we demonstrated the existence of a stable polariton state by achieving strong coupling between cavity photons and magnons in the field-polarised phase of Cu$_2$OSeO$_3$. We also observed dispersive coupling between cavity photons and heli-magnons in the non-collinear phases (helical and conical phases). Cu$_2$OSeO$_3$ is a material system well known to host another exotic topologically protected spin texture state known as magnetic skyrmions \cite{okamura2013microwave}, which stabilise as a skyrmion lattice in a narrow temperature region below the Curie temperature, $T_{\textrm{c}}$. It has been shown that the skyrmion lattice can be excited by magnetic resonance techniques \cite{Onose2012observation, schwarze2015universal}, where the resonance comprises of a collective skyrmion core precession in the counter-clockwise (CCW) and clockwise directions (gyration modes), and the breathing mode where skyrmion cores periodically shrink and enlarge. An open question was left unanswered in our previous work \cite{abdurakhimov2019magnon}, what happens when the cavity photons try to couple with the excitations of a skyrmion lattice? The difficulty in answering this question was mainly due to the high cavity resonance frequency ($\approx 9$ GHz), where a direct interaction with the skyrmion resonance modes ($\approx 1-2$ GHz) could not be observed. 

In this study, we designed two separate cavity systems each optimised to explore the interaction between the cavity and the skyrmion excitation modes. Both cavities are designed to have their primary resonant modes within the expected frequency window of either the CCW or breathing skyrmion modes. We observe the onset of an avoided crossing of the cavity mode, indicating the interaction of the cavity with the skyrmion system by shifting the cavity mode significantly. An enhancement in the cavity linewidth is observed at the degeneracy point, where cavity and skyrmion modes cross. By probing the cavity linewidth as a function of external magnetic field we are able to deduce the effective coupling strength as well as the effective loss rate of the skyrmion excitation. A detailed temperature dependence of the effective coupling strength is extracted for both the CCW and breathing skyrmion modes, indicating the presence of a large number of skyrmions in the vicinity of 57 K. At this temperature the effective coupling strength is largest and we find the cavity-skyrmion hybrid system in the high cooperativity regime.

\section*{Experimental Procedure}

\begin{figure}[t!]
        \includegraphics[width=12cm]{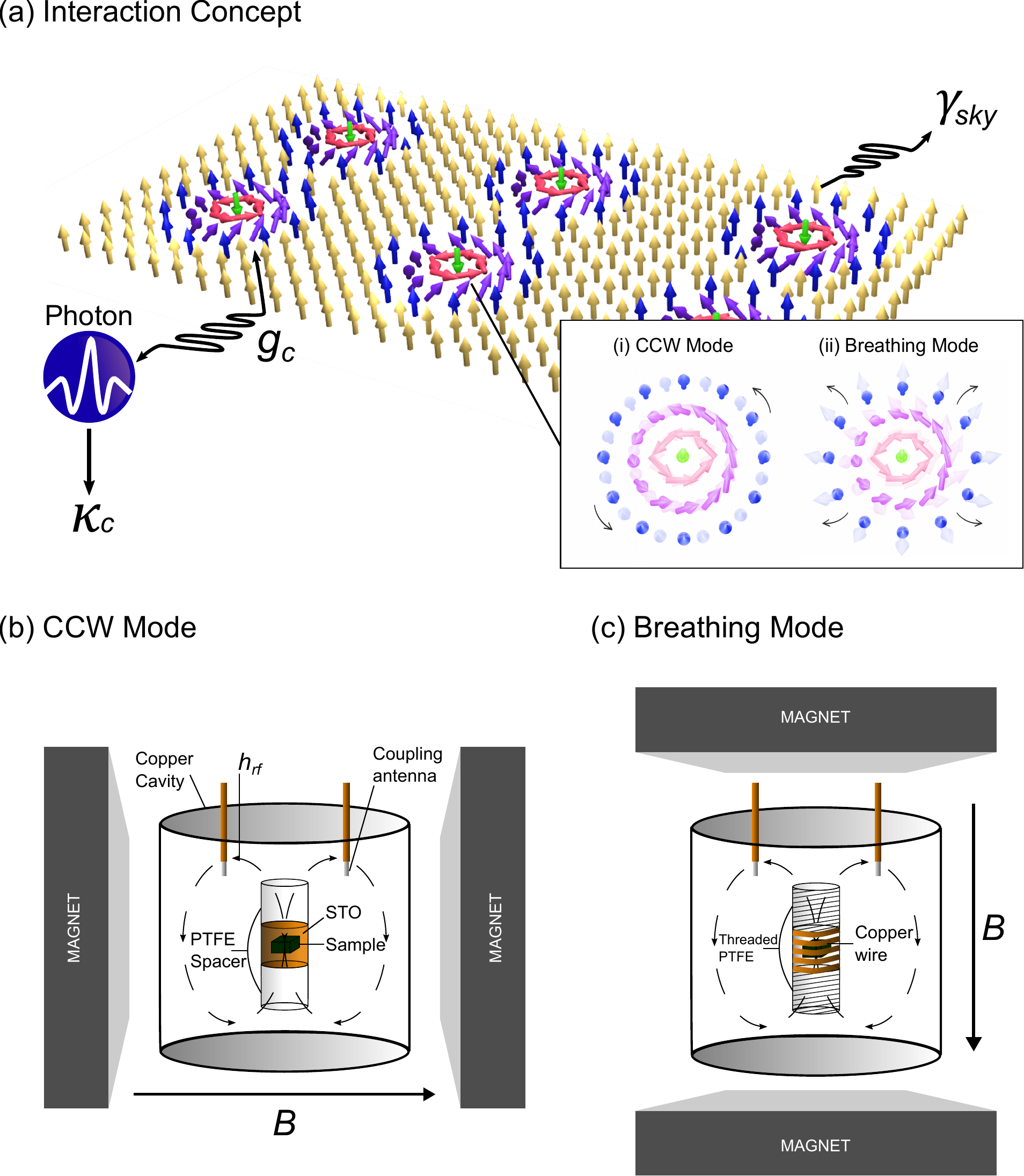}
        \caption{(a) Illustration of the interaction between photons and excitation modes of skyrmion lattice showing the resonant motions of (i) CCW mode where the core of skyrmion rotates, and (ii) breathing mode which involves a periodic oscillation of the skyrmion core diameter. Schematic view of the experimental setup used with the (b) STO cavity where $\textbf{h}_{\textrm{rf}} \perp \mu_0\textbf{H}$ for CCW mode, and (c) helical resonator where $\textbf{h}_{\textrm{rf}} || \mu_0\textbf{H}$ for breathing mode. The cavity and resonator are held in the helium-4 cryostat, and the cryostat sat in between the poles of the electromagnet. A Cu$_2$OSeO$_3$ sample is placed inside the STO cavity and helical resonator with the support of PTFE spacers and threaded PTFE tube, respectively. The coupling to cavity/resonator is achieved by connecting two microwave lines from the vector network analyser (VNA) to the coupling loops in the copper cavity.}
        \label{figure1}
\end{figure}

Figure~\ref{figure1}(a) shows the schematic to highlight the interaction picture between photons (with a loss rate $\kappa_{\textrm{c}}$) and skyrmion resonant modes (with an effective loss rate $\gamma_{\textrm{sky}}$), with $\textrm{g}_{\textrm{c}}$ representing the effective coupling rate between the two excited systems. The dielectric strontium titanate (STO) and helical (helically wrapped copper wire) resonator are used to study the coupling of microwave photons with the counterclockwise (CCW) and breathing skyrmion modes, respectively, in Cu$_2$OSeO$_3$. The remaining of Fig.~\ref{figure1} shows the experimental setup for the (b) STO and (c) helical resonator (see supplementary material for full description \footnote{See Supplemental Materials at [URL will be inserted by
publisher] for detailed experimental method, VNA calibration, cavity measurements outside of skyrmion phase and CPW magnetic resonance experiment.}). The resonance frequency of the primary mode (TE$_{01\delta}$) of the STO cavity, $\omega/2\pi \approx 1.1$ GHz at 57 K is in the expected range of the CCW excitation, as expected from broadband ferromagnetic resonance (FMR) measurements (see supplementary materials) \cite{Onose2012observation}. One can only excite the breathing mode when the microwave magnetic field in the cavity ($\textbf{h}_{\textrm{rf}}$) is parallel to the external magnetic field ($\mu_0\textbf{H}$) \cite{schwarze2015universal}, as shown in Fig.~\ref{figure1}(c). This is achieved by exciting the primary mode of the helical resonator with an excitation frequency, $\omega/2\pi \approx 1.4$ GHz. In the parallel excitation configuration, the excitation of all the other magnetic resonant modes in Cu$_2$OSeO$_3$ should be suppressed, and the resonator mode is expected only to couple to the breathing mode.

\section*{Results $\&$ Discussion}
\subsection*{Onset of an avoided crossing effect in skyrmion phase}

\begin{figure}[h!]
        \includegraphics[width=1\linewidth]{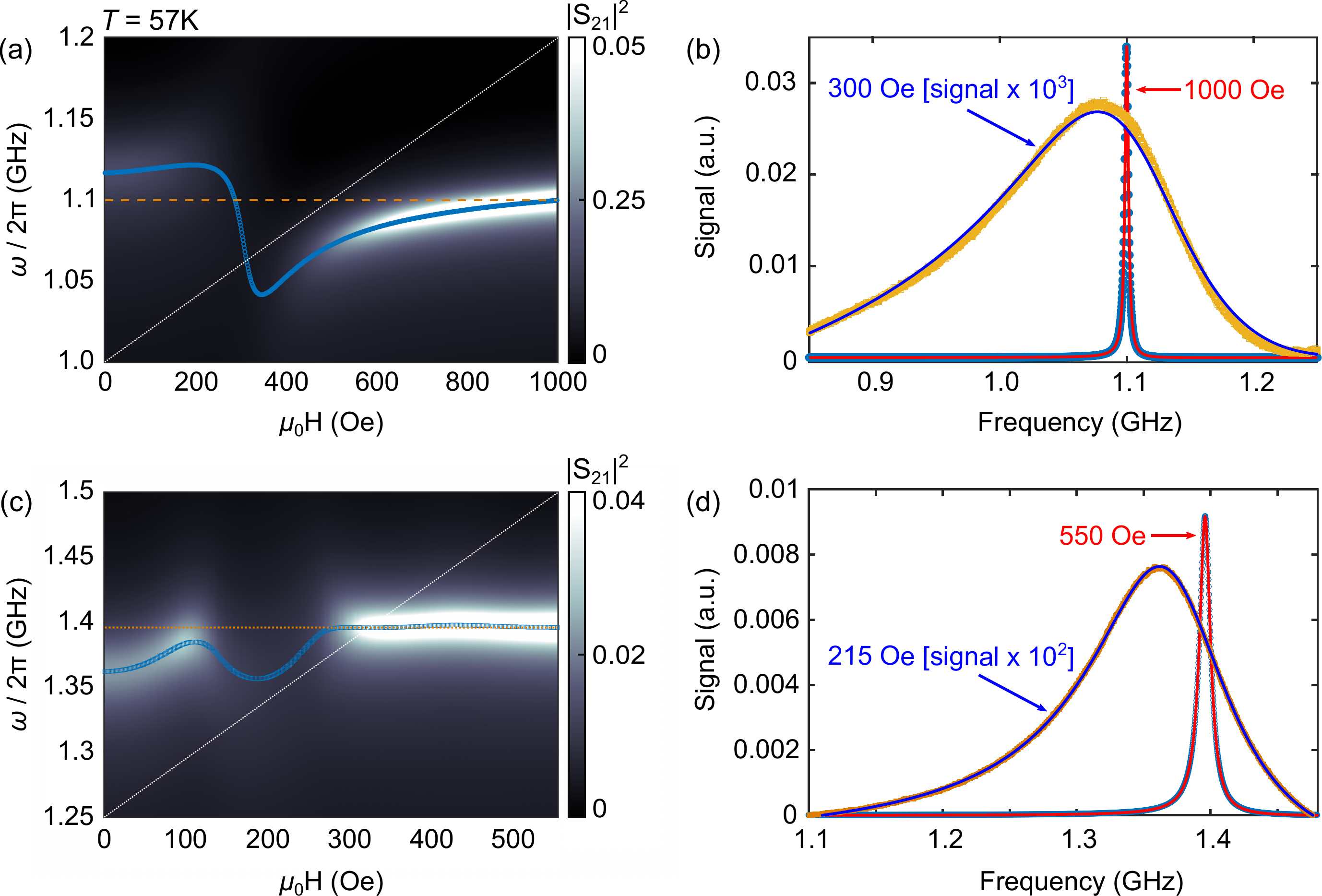}
        \caption{The experimentally measured microwave transmission spectra, $|\textrm{S}_{21}|^2$, at 57 K for (a) CCW mode and (c) breathing mode. The dashed orange line represents the cavity mode, and the resonance peak position plotted as circles (blue). (b) and (d) show line cuts in the frequency domain at high magnetic field away from skyrmion resonance and at field region in the interaction point for CCW and breathing mode, respectively. The data are fitted with a Lorentzian function.}
        \label{figure2}
\end{figure}

Figure~\ref{figure2}(a) shows the microwave transmission, $|\textrm{S}_{21}|^2$, as a function of external magnetic field ($\mu_0$\textbf{H}) and probe frequency for the CCW skyrmion mode at 57 K. The observed spectra exhibit a strong dependence of the cavity mode on $\mu_0$\textbf{H} and a clear shift in the resonance frequency is seen within the field region where a skyrmion lattice is expected. This shift represents an onset of an avoided crossing, due to the coupling of the cavity mode to the skyrmion excitations. A complete avoided crossing, where the cavity resonant mode splits into two peaks in frequency, indicates strong coupling \cite{lachance2019hybrid}. The splitting corresponds to a hybridisation of the two systems, where the separation relates to the coupling strength. The linewidth of the two peaks is defined by the average of the linewidth of the two subsystems. If the linewidth or loss rates of either one or both subsystems is larger than the coupling strength, the two peaks merge. This is the case of the cavity-skyrmion hybrid system we here investigate. In the following, we determine the effective coupling strength and effective skyrmion loss rate to classify the regime of coupling.  

In Fig.~\ref{figure2}(a), $|\textrm{S}_{21}|^2$ drops significantly in the field region between 150~Oe to 400~Oe. To study the evolution of cavity resonance frequency and linewidth in greater detail, we fit a Lorentzian line shape for each magnetic field in the frequency domain \cite{kalarickal2006ferromagnetic}. Figure~\ref{figure2}(b) shows the fitted spectrum at 1000~Oe and in the gap area (300~Oe). The extracted resonance peak position is plotted (blue circles) in Fig.~\ref{figure2}(a). A large increase in linewidth of the cavity mode is visible when Cu$_2$OSeO$_3$ is in the skyrmion phase. Note, that we are observing a single broad lineshape instead of two separated peaks, indicating that the hybrid system exhibits loss rates larger than the coupling strength \cite{bai2016control}.

Figure~\ref{figure2}(c) shows the microwave transmission where the breathing skyrmion mode and the photon mode of the helical resonator interact at 57 K. As in the case of the CCW skyrmion excitation mode, $|\textrm{S}_{21}|^2$ exhibits a decrease in intensity, in a magnetic field space covering the skyrmion phase. The microwave transmission as a function of probe frequency for magnetic fields on and far away from the skyrmion resonance are shown in Fig.~\ref{figure2}(d). By following the same procedure of fitting a Lorenztian function in the frequency domain for each magnetic field, similarly to the CCW case, a clear resonance frequency shift and linewidth broadening is observed. 

It is evident from Fig.~\ref{figure2}(a) and (c), that the photon modes in the STO cavity and the helical resonator couple to the excitation modes of the skyrmion lattice, i.e. CCW gyration and breathing mode, respectively. This is further supported by measurements at temperatures outside of the skyrmion phase ($\textrm{T}<\textrm{T}_{\textrm{sky}}<\textrm{T}_{\textrm{c}}$ and $\textrm{T}>\textrm{T}_{\textrm{c}}$). Here the cavity modes show no signature of coupling in resonance freqeuncy or linewidth (respective colour plots shown in supplementary material \cite{Note1}). 

\subsection*{Temperature dependence of coupling strength}

To identify the epitome of coupling between photons and skyrmion resonance, we analysed the linewidth dependence of the cavity as a function of $\mu_0$\textbf{H} for both CCW and breathing modes. According to the two coupled harmonic oscillators model \cite{huebl2013high, haroche2006exploring,herskind2009realization,abe2011electron, dold2019high},  the cavity linewidth (half-width at half-maximum), $\kappa_{\textrm{(S/H)}}$ (S: STO cavity and H: helical resonator), is given by the following Lorentzian profile, as 

\begin{equation}
    \kappa_{\textrm{(S/H)}} = \kappa_{\textrm{c}} + \frac{\textrm{g}_\textrm{c}^\textrm{2}\gamma_\textrm{sky}}{(\Delta^2 + \gamma_\textrm{sky}^2)}
    \label{equation1}
\end{equation}

where $\kappa_{\textrm{c}}$ represents the loss rate of the cavity photons, $\gamma_\textrm{sky}$ is the effective loss rate for skyrmion resonance, $\textrm{g}_\textrm{c}$ is the effective coupling rate between the photons and skyrmion excitations, and $\Delta$ is the field dependent detuning, given as $\Delta = \frac{\textrm{g}\mu_0\mu_\textrm{B}}{\hbar}$[\textbf{H} $-$ \textbf{H}$_{\textrm{res}}$], with g being the g-factor of the sample, taken as  g $= 2.1$ for Cu$_2$OSeO$_3$ \cite{schwarze2015universal}.

\begin{figure}[h!]
        \includegraphics[width=1\linewidth]{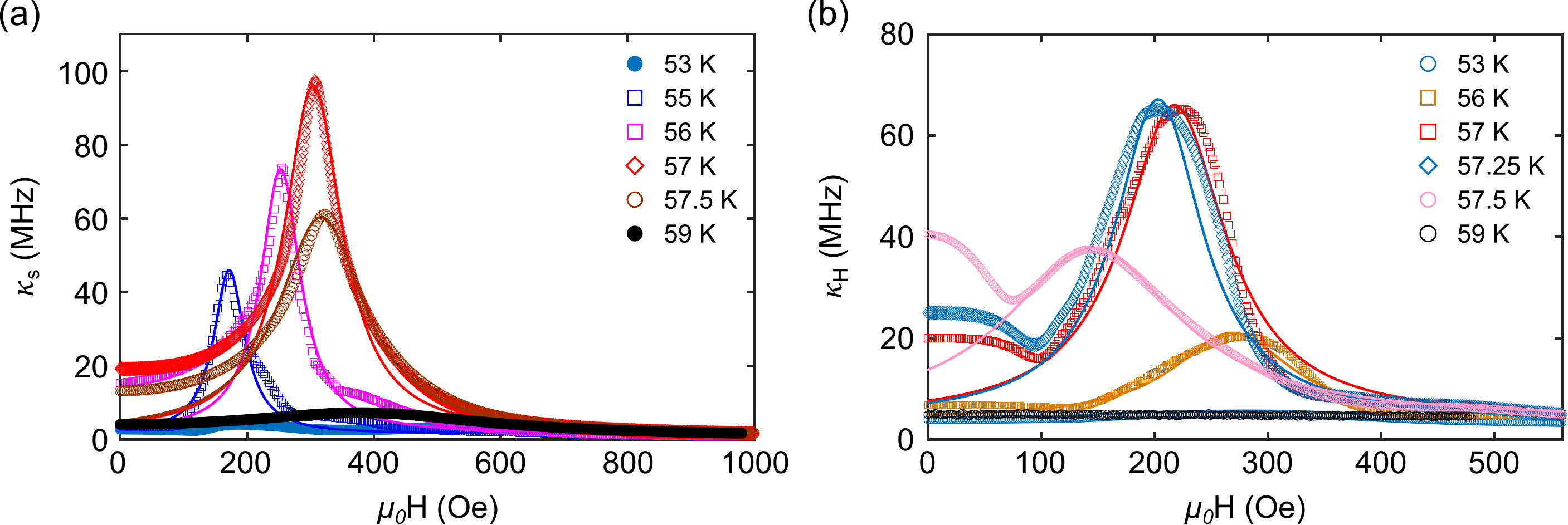}
        \caption{Shows the experimentally extracted linewidth (HWHM) of (a) the STO cavity interacting with CCW mode and (b) helical resonator coupling to the breathing mode as a function of external magnetic field for different temperatures. The data in both is fitted with \autoref{equation1} in the main text.}
        \label{figure3}
\end{figure}

Figures~\ref{figure3}(a) and (b) show the evolution of the extracted linewidth as a function of  $\mu_0$\textbf{H} at different temperatures for the CCW and breathing skyrmion modes, respectively. It is clearly noticeable that in both cases $\kappa_{\textrm{(S/H)}}$ shows a peak like behaviour. At high magnetic fields, the photon mode in the STO cavity and the helical resonator is decoupled from the magnetic system and the measured linewidth mainly remains constant. A large deviation in the linewidth is observed as the magnetic field approaches the expected field region for the skyrmion lattice in the phase diagram at given temperatures. For temperatures outside of the skyrmion phase, i.e. 53 K ($\textrm{T}<\textrm{T}_{\textrm{sky}}$) and 59 K ($\textrm{T}>\textrm{T}_{\textrm{c}}$), we observe no significant change in linewidth as can be seen in Fig.~\ref{figure3}(a) and (b). This provides strong evidence that the photon mode couples to the skyrmion resonance modes.  

In the skyrmion phase, for both CCW and breathing modes, a fit to the data in Figs~\ref{figure3}(a) and (b) is performed using Eq~\ref{equation1} with $\kappa_{\textrm{c}}$, $\gamma_\textrm{sky}$ and $\textrm{g}_\textrm{c}$ as free parameters. The largest effective coupling strength is observed at 57 K for CCW and breathing skyrmion modes. The extracted fitting parameters at 57 K are $\textrm{g}_\textrm{c}^{\textrm{ccw}}/2\pi = 118$ MHz, $\gamma_\textrm{sky}/2\pi = 250$ MHz and $\kappa_{\textrm{c}}^{\textrm{sto}}/2\pi = 1.5$ MHz for the CCW mode, and $\textrm{g}_\textrm{c}^{\textrm{br}}/2\pi = 108$ MHz, $\gamma_\textrm{sky}/2\pi = 309$ MHz and $\kappa_{\textrm{c}}^{\textrm{heli}}/2\pi = 4$ MHz for the breathing mode.

\begin{figure}[h!]
        \includegraphics[width=14 cm]{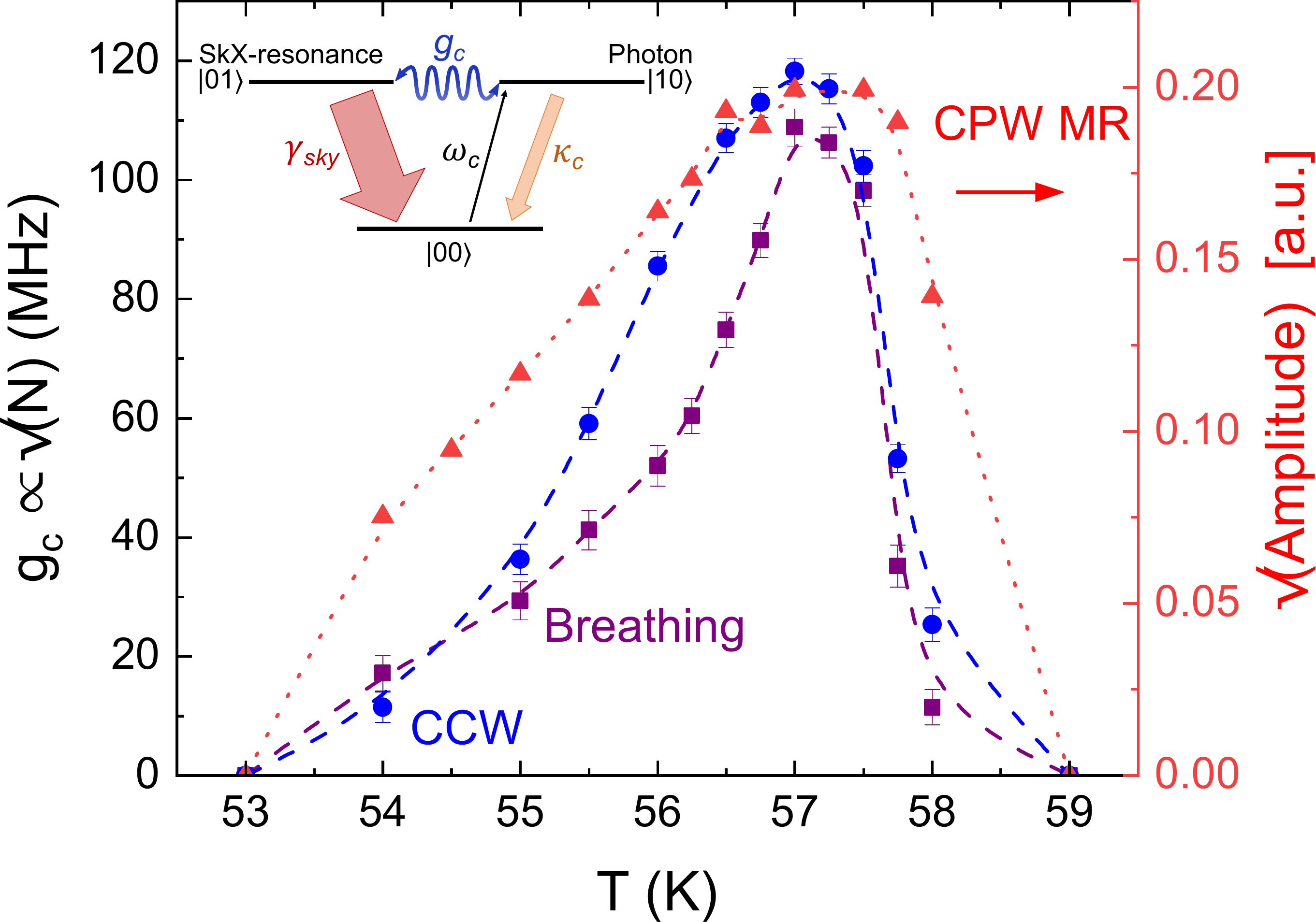}
        \caption{(Left y-axis) Shows the effective coupling strength $\textrm{g}_{\textrm{c}}$ as a function of temperature for CCW mode (blue circles) and breathing mode (purple squares). (Right y-axis) Shows the temperature dependence of the square-root amplitude of resonance peaks extracted from the CPW magnetic resonance experiment. The inset shows a schematic illustrating an interaction between photons and skyrmion excitation. $|00\rangle$ corresponds to the ground state where no photons or skyrmion excitation exists in the cavity, when driven by microwave power, the cavity enters the excited state $|10\rangle$. Coupling of $|10\rangle$ with the excited skyrmion state $|01\rangle$ with coupling strength $\textrm{g}_{\textrm{c}}$ increases the cavity loss rate by having an additional damped channel through skyrmions. The dashed lines are for eye guidance.}
        \label{figure4}
\end{figure}

The type of coupling observed between photons and skyrmion resonance modes falls in a regime where $\kappa_{\textrm{c}} < \textrm{g}_{\textrm{c}} < \gamma_{\textrm{sky}}$ for both CCW and breathing modes. The skyrmion resonance is heavily damped such that the rate of excitation loss happens at a faster rate so that the energy cannot be transferred back and forth between the photon and skyrmion systems. This is illustrated in the schematic shown as an inset in Fig.~\ref{figure4}, where the driven photon system ($|10\rangle$) coupled to the skyrmion resonant mode ($|01\rangle$) opens up an additional path back to the ground state ($|00\rangle$). This results in an enhancement of the linewidth ($\kappa_{\textrm{(S/H)}}$) of the microwave cavity due to its coupling to the skyrmion resonant modes. 

\autoref{figure4} shows the temperature dependence of $\textrm{g}_\textrm{c}$ extracted from the linewidth analysis for both CCW and breathing skyrmion modes. As mentioned earlier, in Cu$_2$OSeO$_3$, the skyrmion phase is expected to exist in a narrow temperature range below $T_{\textrm{c}}$. We see that for both CCW and breathing modes, the highest effective coupling strength is observed for a $~ 1$ K range centered around 57 K. The effective coupling strength gradually falls down when the system is further cooled, and no coupling is observed at 53 K where the linewidth of the STO cavity and the helical resonator becomes unaffected. The experimentally observed trend in the temperature dependence of the effective coupling strength is asymmetric. Above 57.5 K, there is a sharp decrease in $\textrm{g}_\textrm{c}$, reflective of a transition from an ordered magnetic state to the paramagnetic state. 

It is well established that the coupling between photons and an ensemble of spins scales with the square-root of the thermally polarised number of spins  ($N$) involved in the interaction, g$_\textrm{c} \propto \sqrt{N}$ \cite{zollitsch2015high}. We anticipate that for interpreting the results shown in Fig.~\ref{figure4}, one has to think of the coupling  between photons and skyrmion resonance being dependent on the effective number of skyrmions taking part in the interaction. This suggests that around 57 K, a stable hexagonal skyrmion lattice is present, resulting in a greater number of skyrmions. As we move away from this centered region of the  skyrmion phase, the number of skyrmions decrease resulting in a mixed state of magnetic domains and skyrmions. Hence, we see a decrease in the coupling strength between photons and skyrmion resonance. This is further supported by the temperature dependence of the intensity of small-angle neutron scattering (SANS) and the relaxation rate of muon-spin relaxation measurements in Cu$_2$OSeO$_3$ \cite{adams2012long,seki2012formation,hicken2021megahertz}, where the intensity and relaxation rate, respectively, have a maximum around 57 K and a drop was observed at different temperatures in the skyrmion phase. 

In order to further elaborate on the above interpretation, we carried out co-planar waveguide (CPW) magnetic resonance measurements in the skyrmion phase on the same sample used for the cavity experiments (see supplementary material \cite{Note1}). The intensity (I $\propto$ amplitude) of the magnetic resonance peak is proportional to the magnetisation in the system \cite{farle1998ferromagnetic}. In turn, probing the intensity, here mainly governed by the skyrmion population and the effective amplitude of local magnetic moment under thermal fluctuation, in the skyrmion phase of Cu$_2$OSeO$_3$ should be an indicator of the effective number of skyrmions contributing to the magnetic resonance peak amplitude at any given temperature. Therefore, the temperature dependence of the square-root amplitude of the magnetic resonance peaks can reflect the trend seen in the effective coupling strength, i.e. g$_\textrm{c} \propto \sqrt{N}$. \autoref{figure4} (right y-axis) shows the square-root of the amplitude of CPW magnetic resonance peaks as a function of T. The maximum value of the amplitude is observed in the vicinity of 57 K, and is qualitatively similar to the trend observed in g$_\textrm{c}$ for a narrow temperature region where a stable hexagonal skyrmion lattice in Cu$_2$OSeO$_3$ is expected.

In cavity QED, the cooperativity is a dimensionless figure of merit for the coupling strength given as $C = \textrm{g}_\textrm{c}^2/\kappa_{\textrm{c}} \gamma_{\textrm{sky}}$ \cite{huebl2013high}, with $C > 1$ being a condition to achieve coherent coupling. In this work, the largest effective coupling strength value results in $C = 37$ for the coupling of STO cavity photons to the CCW skymion mode and $C = 10$ for the coupling between the helical resonator photons and the breathing skyrmion mode. In both cases, an unstable polariton state is created between photons and skyrmion resonance, coined as a regime of high cooperativity. At resonance nearly all photons inside the cavity are coherently transferred into skyrmion excitations, but transferring the excitations back will require a skyrmion resonance system with a smaller loss rate (i.e. $\gamma_{\textrm{sky}} < \textrm{g}_\textrm{c}$). An earlier theoretical study on the coupling of photons to a breathing skyrmion mode in a thin magnetic disc does conclude that reaching the strong coupling condition in such a system is challenging due to the large loss rate of the skyrmion modes \cite{martinez2019quantum}. 

In conclusion, we have probed the type of coupling between photons in either a STO cavity or a helical resonator to the CCW and breathing skyrmion modes in Cu$_2$OSeO$_3$, respectively. At resonance, an onset of avoided crossing effect is observed, indicating an interaction between the photon mode and skyrmion modes. An enhancement in linewidth at the degeneracy point reveals that the system is in a regime of high cooperativity with $\kappa_{\textrm{c}} < \textrm{g}_{\textrm{c}} < \gamma_{\textrm{sky}}$ for both CCW and breathing modes. The effective coupling strength exhibits a distinct dependence on temperature within the skyrmion phase, and comparing it with the amplitude of CPW magnetic resonance measurements confirms the presence of a skyrmion lattice with a maximum effective number of skyrmions within the vicinity of 57 K. For photon-magnon interactions, the coupling strength scales with the square-root of the number of thermally polarised spins/particles. This study reveals that the interaction between photons and resonant modes of topological spin structures strongly depends on the density of these topological particles instead of the pure spin number in the system. It seems challenging to theoretically predict the number of skyrmions in a bulk system like Cu$_2$OSeO$_3$, with difficulty in an accurate determination of the temperature dependence of internal demagnetisation fields, anisotropy fields, exchange and Dzyaloshinskii-Moriya interactions, and taking into account the quantum effects from spins. A reliable theory support can allow this coupling technique to be used in future as an experimental determination of the number of skyrmions in a given system.

\textit{Note}: In the preparation of this manuscript, authors came to know of the recent pre-print study of dispersive coupling, away from the resonance frequency of skyrmion excitations, between skyrmion modes and cavity photon mode, with a high cooperativity observed at 56.5 K \cite{liensberger2021tunable}.

\section*{Acknowledgements}
The research was supported by JSPS Grants-In-Aid for Scientific Research (Grant Nos. 18H03685, 20H00349, 21H04440), JST PRESTO (Grant No. JPMJPR18L5), Asahi Glass Foundation and JST CREST (Grant No. JPMJCR1874). We would like to thank Mr. Langdon (UCL EEE workshop) for making parts to place the cavity system inside of the cryostat. We are grateful to Prof. Mochizuki for the helpful discussions. 

%

\clearpage

\widetext
\begin{center}
\textbf{\large Supplementary Material: Coupling microwave photons to topological spin-textures in Cu$_2$OSeO$_3$}
\end{center}

\setcounter{equation}{0}
\setcounter{figure}{0}
\setcounter{table}{0}
\setcounter{page}{1}
\makeatletter
\renewcommand{\theequation}{S\arabic{equation}}
\renewcommand{\thefigure}{S\arabic{figure}}
\renewcommand{\bibnumfmt}[1]{[S#1]}
\renewcommand{\citenumfont}[1]{S#1}

\setcounter{equation}{0}
\setcounter{figure}{0}
\setcounter{table}{0}
\setcounter{page}{1}
\makeatletter
\renewcommand{\theequation}{S\arabic{equation}}
\renewcommand{\thefigure}{S\arabic{figure}}
\renewcommand{\bibnumfmt}[1]{[S#1]}
\renewcommand{\citenumfont}[1]{S#1}

\section{C\lowercase{u}$_2$OS\lowercase{e}O$_3$ sample $\&$ Cavity systems}

Single crystals of Cu$_2$OSeO$_3$ were grown by the standard chemical vapor transport method. Microwave transmission spectroscopy technique with a vector network analyser (VNA) was used to measure the scattering parameter, $\textrm{S}_{21}$ as a function of external magnetic field. Strontium titanate (STO) and helical (helically wrapped copper wire) cavities were used to couple photon modes to the CCW and breathing skyrmion modes, respectively, with the schematic of both setups shown in Figure 1 (b) and (c) in the main text. The STO cavity \cite{breeze2016temperature} is a hollow cylinder with dimensions as height $\approx 2.8$ mm, outer diameter $\approx 6.3$ mm and inner diameter $\approx 3$ mm. The helical resonator \cite{batra2017design} is realised by winding the copper wire (wire diameter $\approx 0.5$ mm) on a screw-like threaded PTFE, with a height $\approx 5$ mm, coil diameter $\approx 5$ mm and 5 turns. The cavities (with the sample) were held in a copper cavity (with diameter $\approx 19$ mm and height $\approx 14$ mm) inside of a cryogen free cryostat system, which was placed in between the poles of an electromagnet. Adjustable antennae loops were placed into the sides of the copper cavity (i.e. next to the STO and helical cavities) in order to couple the microwave energy into the cavities \cite{alesini2011power}. Two coaxial cables were used to guide the microwave signal from the two-ports of the VNA to the coupling loops. A full 2-port TOMS calibration (Through-Open-Matched-Short) was performed prior to the experiment. The calibration was done at room-temperature on the whole microwave circuitry in and outside the cryostat, using a R$\&$S ZV-Z129 calibration kit. Raw $\textrm{S}_{21}$ data is shown for both cavity measurements and no background subtraction method is applied.

\section{Cavity photon-mode response outside of Skyrmion phase}
\begin{figure}[h]
        \includegraphics[width=14 cm]{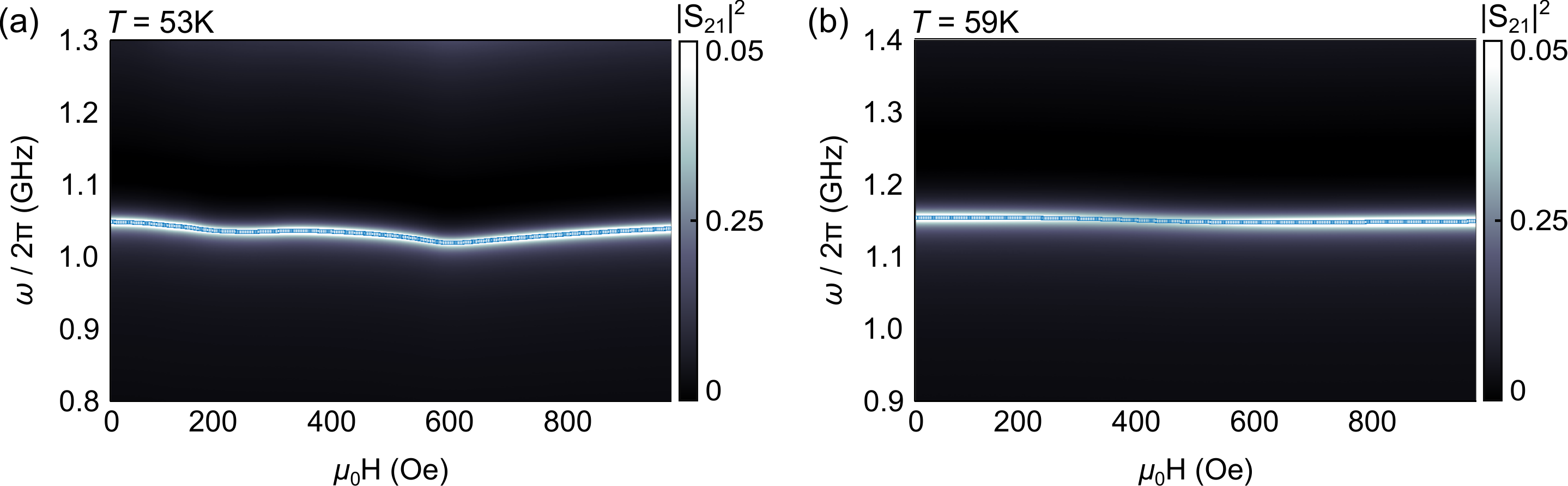}
        \caption{The experimentally measured microwave transmission spectra, $|\textrm{S}_{21}|^2$, for the CCW skyrmion mode at (a) 53~K and (b) 59~K.}
        \label{figure_S1}
\end{figure}

In order to confirm that the cavity photons are interacting with the CCW skyrmion mode, microwave transmission experiments are performed outside of the skyrmion phase temperature region. Figure~\ref{figure_S1}(a) shows the frequency and magnetic field dependence of $|\textrm{S}_{21}|^2$ at 53 K (below the ferrimagnetic Curie temperature and the skyrmion phase), and Figure~\ref{figure_S1}(b) shows the $|\textrm{S}_{21}|^2$ measured at 59 K (above the Tc and in the paramagentic phase). In both cases, the onset of an avoided crossing feature seen in the skyrmion phase (at 57 K in the main text) disappears, and hence, this verifies that the experimentally observed coupling is indeed between the photons and the CCW skyrmion mode.

\begin{figure}[h]
        \includegraphics[width=14 cm]{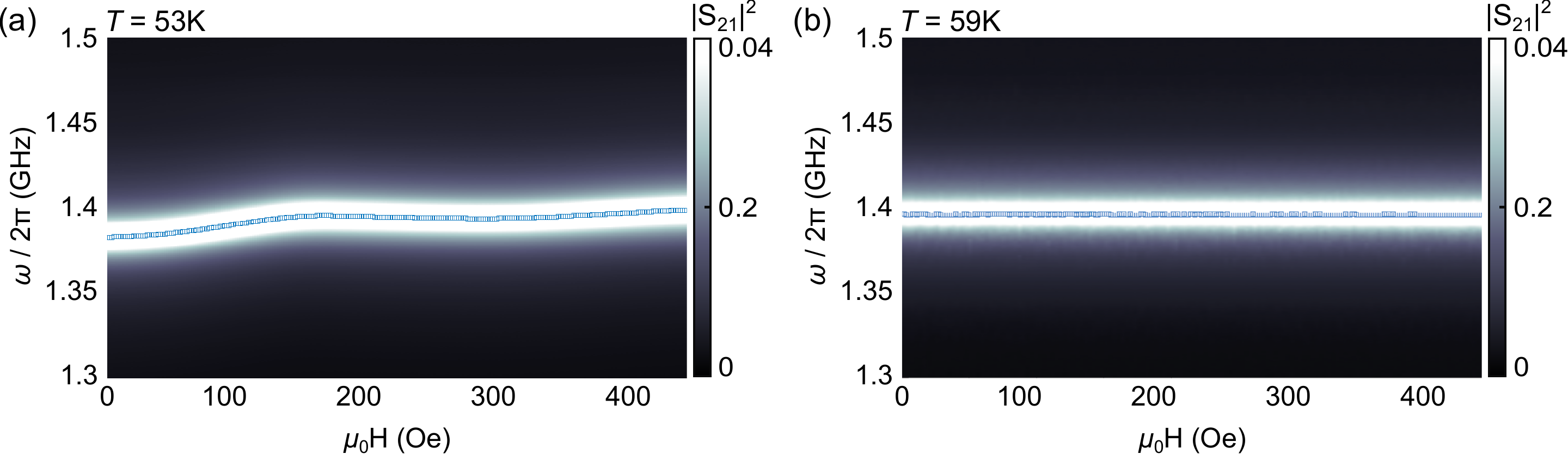}
        \caption{The experimentally measured microwave transmission spectra, $|\textrm{S}_{21}|^2$, for the breathing skyrmion mode at (a) 53~K and (b) 59~K.}
        \label{figure_S2}
\end{figure}

Similar to the CCW case, measurements are carried out for the interaction of the photon mode of the helical cavity and the breathing skyrmion mode, at various other temperatures. Figure~\ref{figure_S2}(a) and (b) shows the transmission signal at 53~K and 59~K, respectively. Here too, no significant deviation in the resonance position of the cavity mode is observed, as the skyrmion lattice is not expected to exist at these temperatures.

\section{Co-planar waveguide magnetic resonance}
\begin{figure}[h!]

\centering
\includegraphics[width=14 cm]{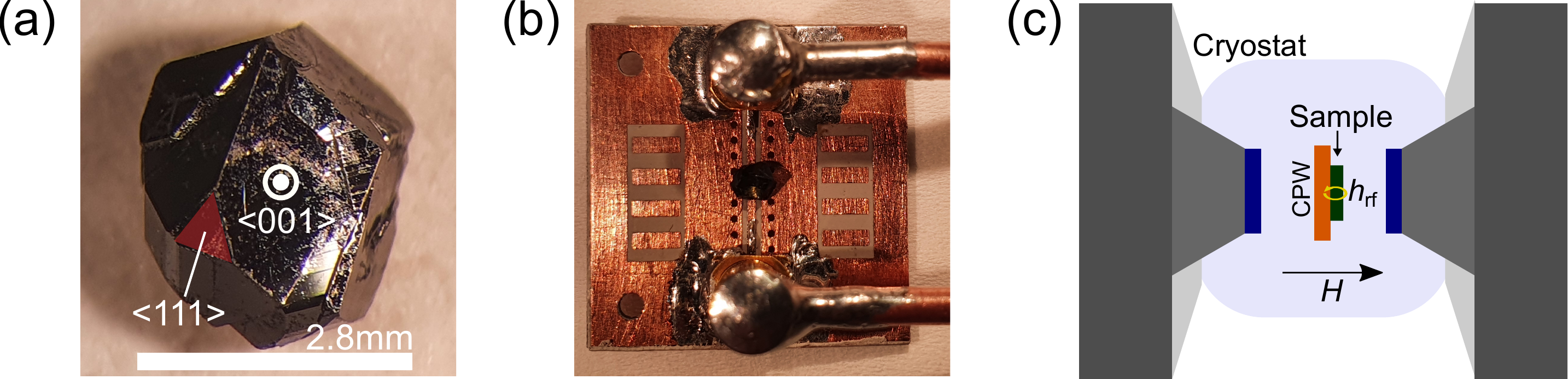}
\caption{\textbf{Experimental setup and sample definition.} (a) Picture of the bulk Cu$_2$OSeO$_3$ sample used. (b) Placement of the sample on the CPW. (c) Typical CPW-FMR experiment schematic. The microwave field is applied perpendicular to the external field from the VNA, where the magnetization dynamics of the sample is excited via the CPW, placed in-between the two electromagnets.}
\label{figure_S3}
        
\end{figure}

A bulk Cu$_2$OSeO$_3$ sample was prepared and placed on a coplanar-waveguide (CPW) as shown in Figs.~\ref{figure_S3}(a) and \ref{figure_S3}(b), respectively. The CPW was then connected to a vector network analyser (VNA) and mounted between the poles of an electromagnet. To excite all magnetic phases of interest, the CPW was placed parallel to the magnet field, yielding a H$_{\textrm{ext}}$ $\perp$ h$_{\textrm{rf}}$ configuration as shown in Fig.~\ref{figure_S3}(c). The data was recorded by varying the static magnetic field at a constant microwave power of 0~dBm and measuring the transmission $S_{21}$ signal.\\

\begin{figure}[h!]
\centering
\includegraphics[width=12 cm]{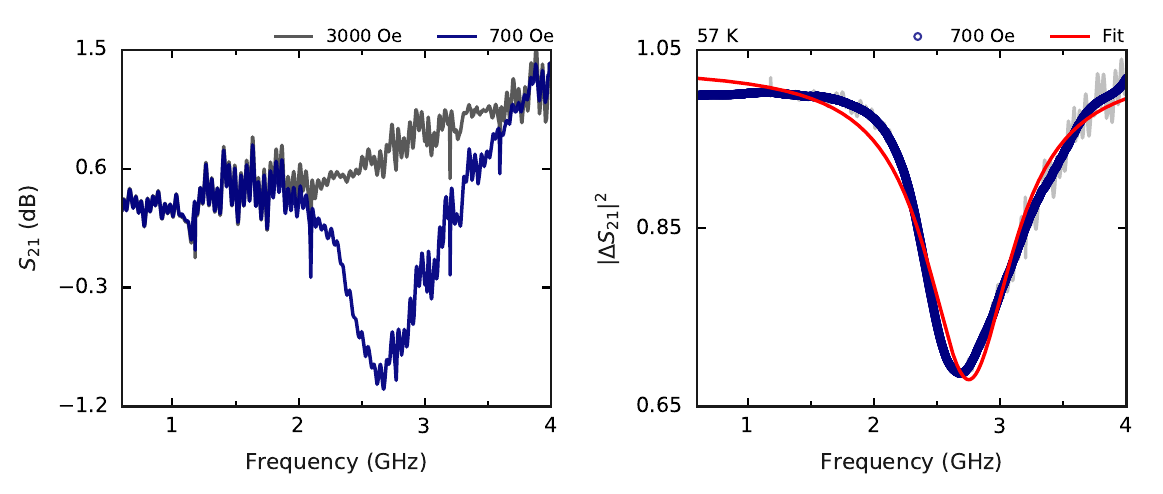}
\caption{\textbf{Processing methods for microwave spectra $S_{21}$.} (Left) Pre-processed transmission spectra $S_{\rm 21}$ at 700 Oe and 3000 Oe at 57~K. (Right) Processed $|S_{\rm 21}|^2$ reflection spectra after background subtraction, numerical filtering and linear conversion.}
\label{figure_S4}
\end{figure}

Fig.~\ref{figure_S4} shows a typical measurement profile of transmission spectra $S_{\rm 21}$ (dB) obtained from the VNA at $\mathrm{\textit T = 57}$~K. The data was processed over three distinctive stages. In the first stage, a high-field background signal has been subtracted from the signal of interest to remove any non-magnetization residuals. Secondly, numerical smoothing utilising a lossless Savitzky-Golay filter~\cite{1964_Savitzky_AnalChem} was employed to improve the signal-to-noise ratio of the obtained signal, resulting $\Delta S_{\rm 21}$ (dB). Lastly, the resulting signal was converted to a linear scale, i.e. $|S_{\rm 21}|^2 = 10^{\Delta S_{\rm 21} /10}$. The same convention is used for all processed data presented in this study.

\begin{figure}[h!]
\centering
\includegraphics[width=14 cm]{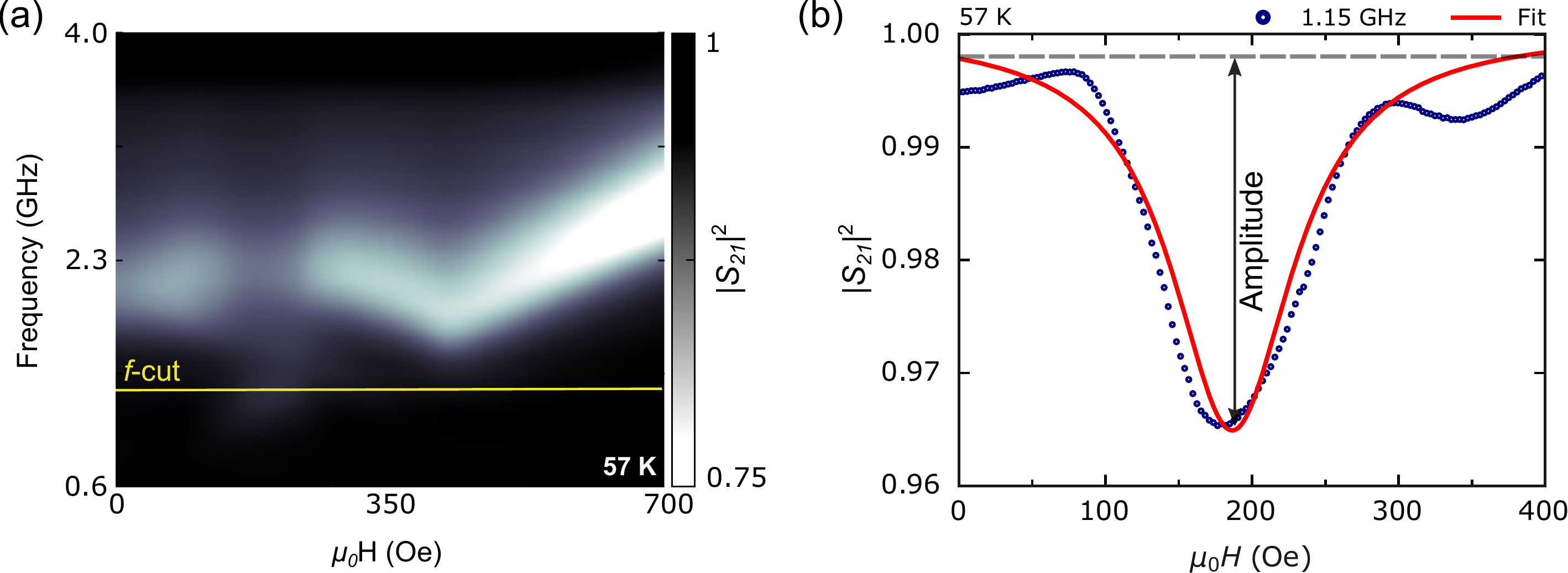}
\caption{\textbf{Define amplitude.} (a) Experimental data for microwave absorption spectra recorded at 57~K. (b) Frequency-cut data plotted at $f=1.15$~GHz.}
\label{figure_S5}
        
\end{figure}

A typical 2D color plot is illustrated in Fig.~\ref{figure_S5}(a). Various magnetization phases including those of helical, counter-clockwise skyrmion (CCW-SkX), conical and field-polarized phases are present as the field is increased. Fig.~\ref{figure_S5}(b) shows a typical frequency-cut spectra at 1.15~GHz. Obtained data was fit by a symmetric-Lorentzian function, where the amplitude was measured as the height of the predominant peak from the baseline shown in gray. Note the baseline is evaluated as the average value between the first and the last data points.

\begin{figure}[h!]
\centering
\includegraphics[width=14 cm]{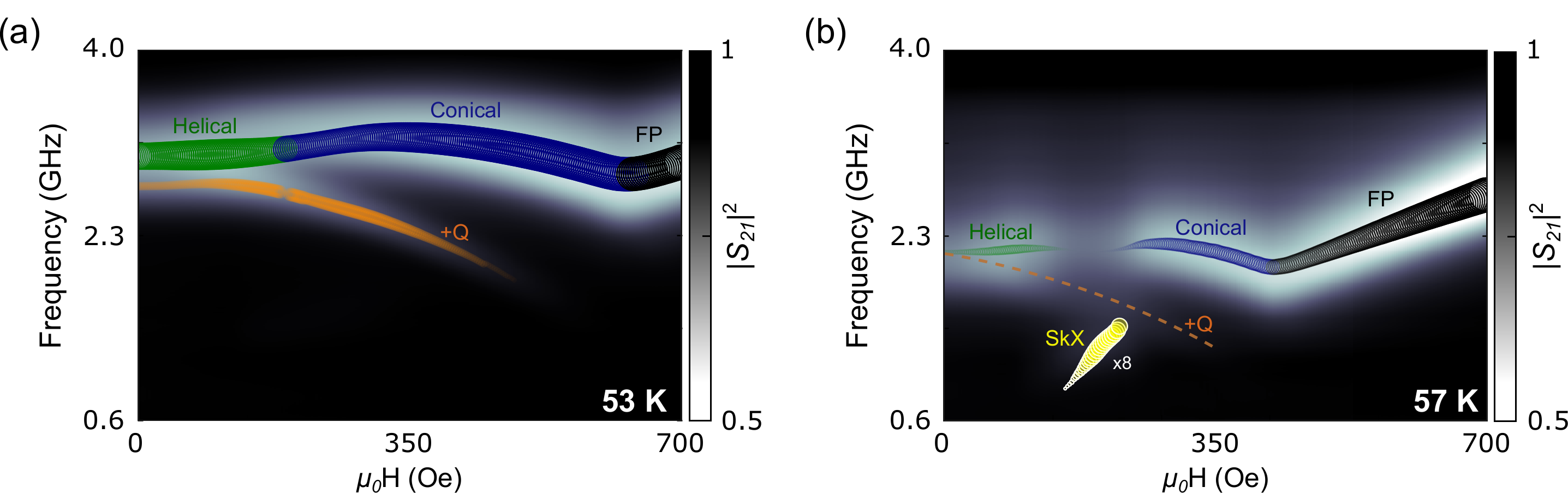}
\caption{\textbf{Frequency-field 2D phase diagrams with weighted amplitude.} (a)~-~(b) Experimental data of microwave absorption spectra recorded respectively at 53~K and 57~K with weighted amplitudes.}
\label{figure_S6}
\end{figure}

2D phase-diagrams with weighted amplitude from the fittings are shown respectively, in Figs.~\ref{figure_S6}(a)~-~(b) at $T = 53~K$ and $57~K$. At 53~K, four prominent magnetization phases can be seen including: helical, conical, +Q branch, and the field-polarized states with an increasing field. On the other hand at 57~K, an obstruct jump is seen between the helical and the conical modes at around 150~-~250~Oe and show clear signatures of CCW-skyrmion peaks at lower frequencies \cite{Onose2012observation,schwarze2015universal}. Increasing the temperature above $T_c$, no FMR response was obtained. 

\end{document}